\documentclass[12pt, a4paper]{article}
\usepackage{mathrsfs}
\usepackage{bbm}
\usepackage{amsmath,amssymb}
\usepackage{color}

\usepackage[titletoc]{appendix}

\allowdisplaybreaks \numberwithin{equation}{section}
\numberwithin{equation}{section} 

 \newtheorem{thm}{Theorem}[section]
 \newtheorem{cor}[thm]{Corollary}
 \newtheorem{lem}[thm]{Lemma}
 \newtheorem{prp}[thm]{Proposition}

 \newtheorem{rmk}[thm]{Remark}
\newenvironment{prf}{\noindent {\it Proof} \ }{\hfill $\Box$}

\newcommand{\eqa}{\begin{eqnarray}}
\newcommand{\eeqa}{\end{eqnarray}}
\newcommand{\beq}{\begin{equation}}
\newcommand{\eeq}{\end{equation}}

\newcommand\pd{\partial} \newcommand\od{\mathrm{d}}

\newcommand{\nn}{\nonumber}

\newcommand\ep{\epsilon}
\newcommand{\ld}{\lambda}
\newcommand{\al}{\alpha}
\newcommand{\gm}{\gamma}

\newcommand{\Gm}{\Gamma}

\newcommand{\dt}{\delta}  \newcommand{\Dt}{\Delta}

\newcommand{\zt}{\zeta}

\newcommand{\diag}{\mathrm{diag}}
\newcommand{\tr}{\mathrm{tr}}

\newcommand{\res}{\mathrm{res}}

\newcommand\C{\mathbb{C}}

\newcommand\Z{\mathbb{Z}}
\newcommand\Zop{\mathbb{Z^{\mathrm{odd}}_+}}

\newcommand\cA{\mathcal{A}}
\newcommand\cD{\mathcal{D}}

\newcommand\cL{\mathcal{L}}
\newcommand\cM{\mathcal{M}}

\newcommand{\set}[1]{\left\{#1\right\}}

\begin{document}
\title{From Additional Symmetries to Linearization of Virasoro Symmetries}
\author{
Chao-Zhong Wu 
\\
\\
{\small SISSA, Via Bonomea 265, 34136 Trieste, Italy}
 }

\date{}
\maketitle

\begin{abstract}
We construct the additional symmetries and derive the
Adler--Shiota--van Moerbeke formula for the two-component BKP
hierarchy. We also show that the Drinfeld--Sokolov hierarchies of
type D, which are reduced from the two-component BKP hierarchy,
possess symmetries written as the action of a series of linear
Virasoro operators on the tau function. It results in that the
Drinfeld--Sokolov hierarchies of type D coincide with Dubrovin and
Zhang's hierarchies associated to the Frobenius manifolds for
Coxeter groups of type D, and that every solution of such a
hierarchy together with the string equation is annihilated by
certain combinations of the Virasoro operators and the time
derivations of the hierarchy.

\vskip 2ex \noindent{\bf Key words}: additional symmetry,
Adler--Shiota--van Moerbeke formula, two-component BKP hierarchy,
Drinfeld--Sokolov
 hierarchy, Virasoro symmetry
\end{abstract}

\section{Introduction}

One of the attractive topics on hierarchies of integrable systems is
to study their additional symmetries. Such symmetries can be
represented in Lax equations with operators depending explicitly on
the time variables as developed by Orlov and Schulman \cite{OS}, and
also in the form of Sato's B\"{a}cklund transformations as the
action of vertex operators on tau function of the hierarchy
\cite{DKJM-KPBKP}. These two strands are connected by a formula of
Adler, Shiota and van Moerbeke \cite{ASvM} in consideration of the
Kadomtsev--Petviashvili (KP) and the two-dimensional Toda lattice
hierarchies aiming at finding the constraints satisfied by matrix
integrals \cite{AvM}. The Adler--Shiota--van Moerbeke (ASvM) formula
for the KP hierarchy was also proved, in a more straightforward way,
by Dickey \cite{Di}.

In this note, we will construct the additional symmetries and derive
the ASvM formula for the two-component BKP hierarchy, which was
first introduced as a bilinear equation by Date, Jimbo, Kashiwara
and Miwa \cite{DJKM} and represented to a Lax form with
pseudo-differential operators in \cite{LWZ}. This hierarchy is found
interesting on aspects such as representation of Lie algebras,
algebraic geometry, topological field theory and
infinite-dimensional Frobenius manifolds \cite{JM, Sh, Ta, WX2}, as
well as for the reductions of it to subhierarchies
\cite{DJKM-reduce, LWZ, Wu}. A direct reduction of the two-component
BKP hierarchy is to the original (one-component) BKP hierarchy
\cite{DKJM-KPBKP}. Recall that the ASvM formula for the BKP
hierarchy was first obtained by van de Leur \cite{vdL} with an
algebraic method, and later proved by Tu \cite{Tu} following the
approach of Dickey \cite{Di}. Our first aim is to extend such
results for the BKP hierarchy to that for the two-component case,
along the line of Dickey and Tu.

Another important reduction of the two-component BKP hierarchy is to
the Drinfeld--Sokolov hierarchy of type $D_n$, that is, the
hierarchy associated to untwisted affine algebra $D_n^{(1)}$ with
the zeroth vertex $c_0$ of the Dynkin diagram marked \cite{DS, LWZ}
(see Section~\ref{sec-virD} for the Lax representation). Based on
this fact, we will derive Virasoro symmetries for the $D_n$-type
Drinfeld--Sokolov hierarchy by using the additional symmetries of
its universal hierarchy. More exactly, let $\tau$ be the tau
function of the Drinfeld--Sokolov hierarchy of type $D_n$ reduced
from that of the two-component BKP hierarchy, then the Virasoro
symmetries for the former hierarchy can be written as
\begin{equation}\label{tausj0}
\frac{\pd\tau}{\pd s_j}=V_j\tau, \quad j\ge-1
\end{equation}
where the operators $V_j$ satisfy
\begin{equation}\label{ViVj0}
[V_i, V_j]=(i-j)V_{i+j}+\dt_{i+j,0}\frac{n}{12}(i^3-i).
\end{equation}
Virasoro symmetries represented in the form \eqref{tausj0} are said
to be linearized.

In a sense inspired by the connection between Gromov--Witten
invariants and integrable systems  \cite{Ko, Wi}, Dubrovin and Zhang
\cite{DZ} proposed a project to classify integrable hierarchies
satisfying certain axioms. These axioms are
\begin{itemize}\setlength{\itemsep}{-0.2ex}
\item[(A1)] the hierarchy
carries a bi-Hamiltonian structure, the dispersionless limit of
which is associated to a semisimple Frobenius manifold \cite{Du};
\item[(A2)] the hierarchy has a tau function defined by the tau-symmetric Hamiltonian densities;
\item[(A3)] linearization of Virasoro symmetries, namely, these symmetries are represented
by the action of certain linear operators $L_j\,(j\ge-1)$ on the tau
function, where the operators $L_j$ 
generate a half Virasoro algebra.
\end{itemize}
For example, both the  Korteweg--de Vries (KdV) and the extended
Toda hierarchies satisfy these axioms, and their tau functions give
generating functions for Gromov--Witten invariants with target
spaces of point and $\C P^1$  \cite{Ko, DZ3, Wi}. Furthermore,
Dubrovin and Zhang conjectured that the Drinfeld--Sokolov
hierarchies of ADE type all satisfy (A1)--(A3). This conjecture has
been confirmed for the $A_n$-type hierarchy (equivalent to the $n$th
KdV or Gelfand--Dickey hierarchy), but is still open for hierarchies
of the other two types. In the case of the Drinfeld--Sokolov
hierarchy of type $D_n$, we will show that the operators $V_j$ are
equal to $L_j$ up to a rescaling of the time variables. Therefore,
in combination with the results in \cite{DLZ, LWZ}, we conclude the
following result.
\begin{thm} \label{thm-main}
The Drinfeld--Sokolov hierarchy of type $D_n$ coincides with the
full hierarchy constructed by Dubrovin and Zhang starting from the
Frobenius manifold on the obit space of the Coxeter group $D_n$.
Moreover, for this hierarchy every analytic solution $\tau$
satisfying the string equation $V_{-1}\tau=\pd\tau/\pd t_1$ admits
the following constraints
\begin{equation}\label{Vc0}
\tilde{L}_j\tau=0, \quad  \tilde{L}_j=V_j-\frac{\pd}{\pd
t_{(2n-2)(j+1)+1}}, \quad j\ge-1.
\end{equation}
Here $t_{(2n-2)(j+1)+1}$ are time variables of the hierarchy and
$\tilde{L}_j$ also obey the communication relation as in
\eqref{ViVj0}.
\end{thm}
We remark that the $D_n$-type Drinfeld--Sokolov hierarchy is
equivalent to several versions of bilinear equations, which are
constructed by Date, Jimbo, Kashiwara and Miwa \cite{DJKM-reduce},
by Kac and Wakimoto \cite{KW, Wu10} in the context of representation
theory, as well as by Givental and Milanov \cite{GM} in singularity
theory.

This paper is arranged as follows. In next section we recall the
definition of the two-component BKP hierarchy. In Sections~3 and 4,
we will write down the additional symmetries and show the ASvM
formula for this hierarchy. By reducing the two-component BKP
hierarchy to the Drinfeld--Sokolov hierarchy of type $D_n$, the
Virasoro symmetries \eqref{tausj0} for this subhierarchy will be
derived in Section~5. In Section~6, we will complete the proof of
theorem~\ref{thm-main}. Finally, Section~7 is devoted to some
remarks.

\section{The two-component BKP hierarchy}
Let us recall the definition of the two-component BKP hierarchy.

Assume $\cA$ is an algebra of smooth functions of a spatial
coordinate $x$, and the derivation $ D=\od/\od x$ induces a
gradation on $\cA$ as
\[
\cA=\prod_{i\geq0}\cA_i, \quad \cA_i\cdot\cA_j\subset\cA_{i+j},
\quad  D(\cA_i)\subset\cA_{i+1}.
\]
Denote $\cD=\set{\sum_{i\in\Z} f_i D^i\mid f_i\in\cA}$ and consider
its two subspaces
\begin{align}\label{}
&\cD^-=\left\{ \sum_{i\in\Z}\sum_{j\ge \max\{0,i-m\}}a_{i,j} D^i
\mid a_{i,j}\in\cA_j, m\in\Z \right\}, \\
&\cD^+=\left\{ \sum_{i\in\Z}\sum_{j\ge \max\{0,m-i\}}a_{i,j} D^i
\mid a_{i,j}\in\cA_j, m\in\Z \right\}.
\end{align}
These subspaces form algebras with multiplication defined by the
following product of monomials
\begin{equation*}
f D^i\cdot g D^j=\sum_{r\geq0}\binom{i}{r}f\, D^r(g)\, D^{i+j-r},
\quad f,g\in\cA.
\end{equation*}
Elements of the algebras $\cD^-$ and $\cD^+$ are called
pseudo-differential operators of the first type and the second type
respectively. Note that in \cite{LWZ} a pseudo-differential operator
of the first type takes the form $\sum_{i<\infty}f_i D^i$ with
$f_i\in\cA$, namely, it only contains finite powers of $D$. Here we
have extended the notion $\cD^-$ slightly, in order to include the
Orlov--Schulman operators that will be introduced in next section.
For any operator $A=\sum_{i\in\Z} f_i D^i\in\cD^\pm$, its
nonnegative part, negative part, residue and adjoint operator are
respectively
\begin{align}\label{Apm}
&A_+=\sum_{i\geq0} f_i  D^i, \quad A_-=\sum_{i<0} f_i  D^i, \quad
\res\,A=f_{-1},\quad A^*=\sum_{i\in\Z}(- D)^i\cdot f_i.
\end{align}

Let
\begin{equation} \label{PhP}
L= D+\sum_{i\ge1}u_i  D^{-i}\in\cD^-, \quad \hat{L}=
D^{-1}\hat{u}_{-1}+\sum_{i\ge1}\hat{u}_i D^i\in\cD^+
\end{equation}
 such that $L^*=- D L D^{-1}$ and $\hat{L}^*=- D\hat{L}
D^{-1}$. The two-component BKP hierarchy is defined by the following
Lax equations:
\begin{align}\label{PPht}
& \frac{\pd L}{\pd t_k}=[(L^k)_+, L], \quad \frac{\pd \hat{L}}{\pd
t_k}=[(L^k)_+, \hat{L}],  \quad
\frac{\pd L}{\pd
\hat{t}_k}=[-(\hat{L}^k)_-, L], \quad \frac{\pd \hat{L}}{\pd
\hat{t}_k}=[-(\hat{L}^k)_-, \hat{L}]
\end{align}
with $k\in\Zop$. Note that $\pd/\pd t_1=\pd/\pd x$; henceforth we
assume $t_1=x$.

One can write the operators $L$ and $\hat{L}$ in a dressing form as
\begin{equation} \label{PPh}
L=\Phi D\Phi^{-1},\quad \hat{L}=\hat{\Phi} D^{-1}\hat{\Phi}^{-1}.
\end{equation}
Here
\begin{align} \label{Phi}
\Phi=1+\sum_{i\ge 1}a_i D^{-i},\quad \hat{\Phi}=1+\sum_{i\ge 1}b_i
D^{i}
\end{align}
are pseudo-differential operators over certain graded algebra that
contains $\cA$, and they satisfy
\begin{equation}\label{phipsi}
\Phi^*= D\Phi^{-1} D^{-1},\quad \hat{\Phi}^*= D\hat{\Phi}^{-1}
D^{-1}.
\end{equation}
Given $L$ and $\hat{L}$, the dressing operators $\Phi$ and
$\hat{\Phi}$ are determined uniquely up to a multiplication to the
right by operators of the form \eqref{Phi} and \eqref{phipsi} with
constant coefficients. The two-component BKP hierarchy \eqref{PPht}
can be redefined as
\begin{align}
&\frac{\pd \Phi}{\pd t_k}=- (L^k)_-\Phi, \quad
\frac{\pd \hat{\Phi}}{\pd t_k}=\bigl((L^k)_+ -\dt_{k1} \hat{L}^{-1}\bigr)\hat{\Phi}, \label{ppt1}\\
&\frac{\pd \Phi}{\pd \hat{t}_k}=- (\hat{L}^k)_-\Phi, \quad \frac{\pd
\hat{\Phi}}{\pd \hat{t}_k}=(\hat{L}^k)_+\hat{\Phi} \label{ppt2}
\end{align}
with $k\in\Zop$.

Denote $\mathbf{t}=(t_1,t_3,t_5,\dots)$ and
$\hat{\mathbf{t}}=(\hat{t}_1,\hat{t}_3,\hat{t}_5,\dots)$. Introduce
two wave functions
\begin{align}\label{wavef}
w(z)=w(\mathbf{t}, \hat{\mathbf{t}}; z)=\Phi e^{\xi(\mathbf{t};z)},
\quad \hat{w}(z)=\hat{w}(\mathbf{t}, \hat{\mathbf{t}}; z)=\hat{\Phi}
e^{x z+\xi(\hat{\mathbf{t}};-z^{-1})},
\end{align}
where the function $\xi$ is defined as $\xi(\mathbf{t};
z)=\sum_{k\in\Zop} t_k z^k$, and $D^i e^{x z}=z^i e^{x z}$ for any
$i\in\Z$. It is easy to see
\[
L\,w(z)=z w(z), \quad \hat{L} \hat{w}(z) = z^{-1} \hat{w}(z).
\]

The hierarchy \eqref{PPht} carries infinitely many bi-Hamiltonian
structures \cite{Wu, WX}. The Hamiltonian densities, which are
proportional to the residues of $L^k$ and $\hat{L}^k$, satisfy the
tau-symmetric condition. Hence given a solution of the two-component
BKP hierarchy, there locally exists a tau function
$\tau(\mathbf{t},\hat{\mathbf{t}})$ such that
\begin{equation}\label{tau-1}
\od\left(2\,\pd_x\,\log\tau\right)=\sum_{k\in\Zop}
\left(\res\,L^k\,\od t_k+\res\,\hat{L}^k\,\od\hat{t}_k\right).
\end{equation}
With this tau function the wave functions can be written as
\begin{align}\label{wtau}
w(\mathbf{t},\hat{\mathbf{t}};z)=\frac{\tau(\mathbf{t}-2[z^{-1}],
\hat{\mathbf{t}})}{\tau(\mathbf{t},\hat{\mathbf{t}})}
e^{\xi(\mathbf{t};z)}, \quad \hat{w}(\mathbf{t},\hat{\mathbf{t}};z)
=\frac{\tau(\mathbf{t},\hat{\mathbf{t}}+2[z])}{\tau(\mathbf{t},\hat{\mathbf{t}})}
e^{\xi(\hat{\mathbf{t}};-z^{-1})}
\end{align}
where $[z]=\left(z,z^3/3,z^5/5,\dots\right)$.
\begin{thm}[\cite{LWZ}]\label{thm-bl}
The hierarchy defined by \eqref{ppt1} and \eqref{ppt2} is equivalent
to the following bilinear equation
\begin{align}\label{bl2BKP}
&-\res_{z=\infty}  z^{-1} \tau(\mathbf{t}-2[z^{-1}],
\hat{\mathbf{t}})\tau(\mathbf{t}'+2[z^{-1}],
\hat{\mathbf{t}}')e^{\xi(\mathbf{t}-\mathbf{t}';z)}\od z \nn \\
 =&
\res_{z=0} z^{-1} \tau(\mathbf{t},
\hat{\mathbf{t}}+2[z])\tau(\mathbf{t}',
\hat{\mathbf{t}}'-2[z])e^{\xi(\hat{\mathbf{t}}-\hat{\mathbf{t}}';-z^{-1})}\od
z.
\end{align}
\end{thm}
The proof of this theorem is mainly based on the following lemma.
\begin{lem}[see, for example, \cite{DKJM-KPBKP}] \label{thm-PQz}
For any pseudo-differential operators $P$ and $Q$ of the same type,
it holds that
\begin{equation*}\label{}
\res_z(P e^{z x}\cdot Q^* e^{-z x})=\res (P Q).
\end{equation*}
\end{lem}
Here and below the residue of a Laurent series is defined as
$\res_z\sum_i f_i z^i=f_{-1}$.

In \eqref{PPht} if $\hat{L}$ vanishes, then one obtains the BKP
hierarchy, of which the bilinear equation is \eqref{bl2BKP} with the
right hand side replaced by $1$, see \cite{DKJM-KPBKP}. The
following fact is observed from the bilinear equation
\eqref{bl2BKP}.
\begin{prp}
When either $\hat{\mathbf{t}}$ or $\mathbf{t}$ is fixed, the
two-component BKP hierarchy is reduced to the BKP hierarchy with
time variables $\mathbf{t}$ or
 $\hat{\mathbf{t}}$ respectively.
\end{prp}

\section{Orlov--Schulman operators and additional symmetries}

We are to construct additional symmetries for the two-component BKP
hierarchy by using the Orlov--Schulman operators, the coefficients
of which may depend explicitly on the time variables of the
hierarchy.

With the dressing operators given in \eqref{PPh}, we introduce
\begin{equation*}\label{}
M=\Phi\Gm\Phi^{-1}, \quad
\hat{M}=\hat{\Phi}\hat{\Gm}\hat{\Phi}^{-1},
\end{equation*}
where
\[
\Gm=\sum_{k\in\Zop}k t_k  D^{k-1}, \quad \hat{\Gm}=x+\sum_{k\in\Zop}
k\hat{t}_k D^{-k-1}.
\]
If we assume the degrees of $t_k$ and $\hat{t}_k$ are equal to $k$,
then $M$ and $\hat{M}$ are pseudo-differential operators of the
first and the second types respectively.

It is easy to see the following
\begin{lem}\label{thm-Mw}
The operators $M$ and $\hat{M}$ satisfy
\\
(1)
\begin{equation}\label{}
[L, M]=1, \quad [\hat{L}^{-1},\hat{M}]=1;
\end{equation}
(2)
\begin{equation}\label{}
M w(z)=\pd_z w(z), \quad \hat{M} \hat{w}(z)=\pd_z \hat{w}(z);
\end{equation}
(3)
 for $\dot{M}=M$ or $\hat{M}$,
\begin{equation}\label{Mt}
\frac{\pd \dot{M}}{\pd t_k}=[(L^k)_+,\dot{M}],\quad
 \frac{\pd \dot{M}}{\pd \hat{t}_k}=[-(\hat{L}^k)_-, \dot{M}].
\end{equation}
\end{lem}

Given any pair of integers $(m,l)$ with $m\ge0$, let
\begin{equation}\label{}
 A_{m l}=M^m
L^l-(-1)^l L^{l-1}M^m L, \quad  \hat{A}_{m l}=\hat{M}^m
\hat{L}^{-l}-(-1)^l \hat{L}^{-l+1}\hat{M}^m \hat{L}^{-1}.
\end{equation}
It is easy to check
\begin{equation}\label{Astar}
A_{m l}^*=-D  A_{m l} D^{-1}, \quad \hat{A}_{m l}^*=-D \hat{A}_{m l}
 D^{-1}.
\end{equation}
The following equations are well defined
\begin{align}\label{}
&\frac{\pd \Phi}{\pd s_{m l}}=- (A_{m l})_-\Phi,
\quad \frac{\pd \hat{\Phi}}{\pd s_{m l}}=(A_{m l})_+\hat{\Phi}, \label{sml}\\
&\frac{\pd \Phi}{\pd \hat{s}_{m l}}=- (\hat{A}_{m l})_-\Phi, \quad
\frac{\pd \hat{\Phi}}{\pd \hat{s}_{m l}}=(\hat{A}_{m
l})_+\hat{\Phi}. \label{shml}
\end{align}
We assume that these flows commute with $\pd/\pd x$.

The following lemma is clear.
\begin{lem}\label{thm-MLs}
The flows \eqref{sml} and \eqref{shml} satisfy
\begin{align*}
&\frac{\pd L}{\pd \dot{s}_{m l}}=[-(\dot{A}_{m l})_-, L], \quad
\frac{\pd \hat{L}}{\pd \dot{s}_{m l}}=[(\dot{A}_{m l})_+, \hat{L}],
\\
&\frac{\pd M}{\pd \dot{s}_{m l}}=[-(\dot{A}_{m l})_-, M], \quad
\frac{\pd \hat{M}}{\pd \dot{s}_{m l}}=[(\dot{A}_{m l})_+, \hat{M}],
\\
&\frac{\pd w(z)}{\pd \dot{s}_{m l}}=-(\dot{A}_{m l})_-w(z), \quad
\frac{\pd \hat{w}(z)}{\pd \dot{s}_{m l}}=(\dot{A}_{m
l})_+\hat{w}(z),
\end{align*}
where $\dot{s}_{m l}=s_{m l}, \hat{s}_{m l}$ correspond to
$\dot{A}_{m l}=A_{m l}, \hat{A}_{m l}$ respectively.
\end{lem}

\begin{prp}\label{thm-st}
The flows \eqref{sml} and \eqref{shml} commute with those in
\eqref{ppt1} and \eqref{ppt2} that compose the  two-component BKP
hierarchy. Namely, for any $\dot{s}_{m l}=s_{m l}, \hat{s}_{m l}$
and $\bar{t}_k=t_k, \hat{t}_k$ one has
\begin{equation}\label{st}
\left[\frac{\pd}{\pd \dot{s}_{m l}}, \frac{\pd}{\pd
\bar{t}_k}\right]=0, \quad m\ge0, ~~ l\in\Z, ~~ k\in\Zop.
\end{equation}
\end{prp}
\begin{prf}
The  proposition is checked case by case with the help of
Lemmas~\ref{thm-Mw} and \ref{thm-MLs}.
For example,
\begin{align}\label{}
&\left[\frac{\pd}{\pd \hat{s}_{m l}}, \frac{\pd}{\pd
t_k}\right]\hat{\Psi} \nn\\
=& [ (L^k)_+-\dt_{k1}\hat{L}^{-1},(\hat{A}_{m l})_+]\hat{\Psi}
+\left([-(\hat{A}_{m l})_-, L^k]_+ - \dt_{k1}[(\hat{A}_{m l})_+,
\hat{L}^{-1}]\right)\hat{\Psi} \nn \\
&-[(L^k)_+,\hat{A}_{m l}]_+\hat{\Psi} =0. \nn
\end{align}
The proposition is proved.
\end{prf}

Proposition~\ref{thm-st} implies that the flows \eqref{sml} and
\eqref{shml} give symmetries for the two-component BKP hierarchy.
Such symmetries are called the \emph{additional symmetries}.

\begin{rmk}
The vector fields $\pd/\pd s_{0,2 i+1}$ (resp. $\pd/\pd\hat{s}_{0,-2
i-1}$) cannot be identified to $2\pd/\pd t_{2 i+1}$ (resp.
$2\pd/\pd\hat{t}_{2 i+1}$). In fact, they act differently on either
$M$ or $\hat{M}$.
\end{rmk}

Now we compute the commutation relation between the additional
symmetries. Observe that the commutator $X=[A_{m l}, A_{m' l'}]$ is
a polynomial in $M$ and $L^{\pm1}$, and it satisfies $X^*=-D X
D^{-1}$, hence there exist constants $c_{m l, m' l'}^{q r}$ such
that
\[
[A_{m l}, A_{m' l'}]=\sum_{q,r}c_{m l, m' l'}^{q r}A_{q r}.
\]
In the same way, one also has $[\hat{A}_{m l}, \hat{A}_{m'
l'}]=\sum_{q,r}c_{m l, m' l'}^{q r}\hat{A}_{q r}$. For example,
\[
c_{0l,0l'}^{qr}=0, \quad  c_{1,2 i+1;1,2 j+1}^{q
r}=4(i-j)\dt_{q1}\dt_{r,2(i+j)+1}.
\]
\begin{prp}\label{thm-ssh}
Acting on the dressing operators $\Phi$ and $\hat{\Phi}$ (or the
wave functions $w$ and $\hat{w}$), the vector fields of the
additional symmetries \eqref{sml} and \eqref{shml} satisfy
\begin{align}
&\left[\frac{\pd}{\pd s_{m l}}, \frac{\pd}{\pd s_{m'
l'}}\right]=-\sum_{q,r}c_{m l, m' l'}^{q r}\frac{\pd}{\pd s_{q r}},
\\
& \left[\frac{\pd}{\pd \hat{s}_{m l}}, \frac{\pd}{\pd \hat{s}_{m'
l'}}\right]=\sum_{q,r}c_{m l, m' l'}^{q r}\frac{\pd}{\pd \hat{s}_{q r}},\\
 &\left[\frac{\pd}{\pd s_{m l}}, \frac{\pd}{\pd \hat{s}_{m'
l'}}\right]=0
\end{align}
 It means that these vector
fields generate a $w_{\infty}^B\times w_{\infty}^B$-algebra (cf.,
for example, \cite{Tu}).
\end{prp}
\begin{prf}
The conclusion follows from a straightforward calculation.
\end{prf}

To prepare for the next section, we introduce two generating
functions of operators as
\begin{align}\label{Y}
&Y(\ld,
\mu)=-\sum_{m=0}^\infty\frac{(\mu-\ld)^m}{m!}\sum_{l=-\infty}^\infty
\ld^{-m-l}(A_{m,m+l})_-,
\\
&\hat{Y}(\ld,
\mu)=\sum_{m=0}^\infty\frac{(\mu-\ld)^m}{m!}\sum_{l=-\infty}^\infty
\ld^{-m-l}(\hat{A}_{m,m+l})_+ \label{Yh}
\end{align}
with parameters $\ld$ and $\mu$.
\begin{lem}\label{thm-Yw}
The action of the generators \eqref{Y} and \eqref{Yh} on the wave
functions \eqref{wavef} reads
\begin{align}\label{}
2
Y(\ld,\mu)w(z)=&-w(-\ld)\pd_x^{-1}\left(w_x(\mu)w(z)-w(\mu)w_x(z)\right)
\nn\\
&+w(\mu)\pd_x^{-1}\left(w_x(-\ld)w(z)-w(-\ld)w_x(z)\right), \label{Yw} \\
2 \hat{Y}(\ld,\mu)\hat{w}(z)=&\hat{w}(-\ld)\pd_{\hat{x}}^{-1}
\left(\hat{w}_{\hat{x}}(\mu)\hat{w}(z)-\hat{w}(\mu)\hat{w}_{\hat{x}}(z)\right)\nn\\
&-\hat{w}(\mu)\pd_{\hat{x}}^{-1}
\left(\hat{w}_{\hat{x}}(-\ld)\hat{w}(z)-\hat{w}(-\ld)\hat{w}_{\hat{x}}(z)\right)
\label{Yhwh}
\end{align}
with ${\hat{x}}=\hat{t}_1$. Here the subscripts $x$ and $\hat{x}$
mean the derivatives with respect to them.
\end{lem}
\begin{prf}
Let us check the second equality first. The property \eqref{Astar}
implies $(\hat{A}_{m,m+l})_+(1)=0$, then using Lemmas~\ref{thm-PQz}
and \ref{thm-Mw} we have
\begin{align*}
&(\hat{A}_{m,m+l})_+\hat{w}(z) \\
=&\sum_{i\ge1}\res\left(\hat{A}_{m,m+l} D^{-i-1}\right)D^i\hat{w}(z)
\\
=&\sum_{i\ge1}\Big(\res_\zt\left(\hat{M}^m\hat{\Phi}D^{m+l}e^{x\zt+\xi(\hat{\bf{t}};-\zt^{-1})}
\cdot(\hat{\Phi}^{-1}D^{-i-1})^*e^{-x\zt-\xi(\hat{\bf{t}};-\zt^{-1})}\right)
\\
&-(-1)^{m+l}\res_\zt\left(\hat{L}^{-m-l+1}\hat{M}^m\hat{\Phi}D
e^{x\zt+\xi(\hat{\bf{t}};-\zt^{-1})}
\cdot(\hat{\Phi}^{-1}D^{-i-1})^*e^{-x\zt-\xi(\hat{\bf{t}};-\zt^{-1})}\right)\Big)D^i\hat{w}(z)
\\
=&\sum_{i\ge1}\Big(\res_\zt\left(\zt^{m+l}\pd_\zt^m\hat{w}(\zt)
\cdot(-D)^{-i}\zt^{-1}\hat{w}(-\zt)\right)
\\
&-(-1)^{m+l}\res_\zt\left(\pd_\zt^m\zt^{m+l-1}\hat{w}(\zt)
\cdot(-D)^{-i}\hat{w}(-\zt)\right)\Big)D^i\hat{w}(z)
\\
=&-\res_\zt\left(\zt^{m+l-1}\pd_\zt^m\hat{w}(\zt)\cdot
D^{-1}(\hat{w}(-\zt)\hat{w}_x(z))\right)
\\
&+(-1)^{m+l}\res_\zt\left(\pd_\zt^m(\zt^{m+l-1}\hat{w}(\zt)) \cdot
D^{-1}(\hat{w}(-\zt)\hat{w}_x(z))\right).
\end{align*}
Hence
\begin{align*}
&\hat{Y}(\ld,\mu)\hat{w}(z)
\\
=&\sum_{m=0}^\infty\frac{(\mu-\ld)^m}{m!}\sum_{l=-\infty}^\infty
\ld^{-m-l}\Big(-\res_\zt\left(\zt^{m+l-1}\pd_\zt^m\hat{w}(\zt)\cdot
D^{-1}(\hat{w}(-\zt)\hat{w}_x(z))\right)
\\
&+(-1)^{m+l}\res_\zt\left(\pd_\zt^m(\zt^{m+l-1}\hat{w}(\zt)) \cdot
D^{-1}(\hat{w}(-\zt)\hat{w}_x(z))\right)\Big)
\\
=&-\sum_{m=0}^\infty\frac{(\mu-\ld)^m}{m!}\pd_\ld^m\hat{w}(\ld)\cdot
D^{-1}(\hat{w}(-\ld)\hat{w}_x(z))
\\
&+\sum_{l'=-\infty}^\infty (-\ld)^{-l'}\res_\zt\left(
(\zt+\mu-\ld)^{l'-1}\hat{w}(\zt+\mu-\ld)D^{-1}(\hat{w}(-\zt)\hat{w}_x(z))\right)
\\
=&-\hat{w}(\mu)D^{-1}(\hat{w}(-\ld)\hat{w}_x(z))+\hat{w}(-\ld)D^{-1}(\hat{w}(\mu)\hat{w}_x(z)).
\end{align*}
It leads to
\begin{align}\label{}
2 \hat{Y}(\ld,\mu)\hat{w}(z)=&\hat{w}(-\ld)D^{-1}
\left(\hat{w}(\mu)\hat{w}_{x}(z)-\hat{w}_{x}(\mu)\hat{w}(z)\right)\nn\\
&-\hat{w}(\mu)D^{-1}
\left(\hat{w}(-\ld)\hat{w}_{x}(z)-\hat{w}_{x}(-\ld)\hat{w}(z)\right).
\label{Yhwh2}
\end{align}
To rewrite this equality as \eqref{Yhwh}, we denote
$\rho=\res\hat{L}$ and recall $
\pd_{\hat{x}}\hat{w}(z)=-\hat{L}_-\hat{w}(z)=-D^{-1}(\rho
\hat{w}(z))$. Since
\begin{align*}
&\pd_{\hat{x}}D^{-1}
\left(\hat{w}(\mu)\hat{w}_{x}(z)-\hat{w}_{x}(\mu)\hat{w}(z)\right)
\\
=&D^{-1} \left(-D^{-1}(\rho\hat{w}(\mu))\hat{w}_{x}(z)+
\hat{w}_{x}(\mu)D^{-1}(\rho\hat{w}(z))\right)
\\
=&-D^{-1}(\rho\hat{w}(\mu))\hat{w}(z)+
\hat{w}(\mu)D^{-1}(\rho\hat{w}(z))
\\
=&\hat{w}_{\hat{x}}(\mu)\hat{w}(z)-
\hat{w}(\mu)\hat{w}_{\hat{x}}(z),
\end{align*}
then \eqref{Yhwh2} is recast to \eqref{Yhwh}. The verification of
the equality \eqref{Yw} is easier. The lemma is proved.
\end{prf}

\begin{rmk}
Suppose $\hat{\bf{t}}$ is fixed, then $w(z)$ is a wave function of
the BKP hierarchy. In this case, the equality \eqref{Yw} is just
equation~(24) in \cite{Tu}, with $Y(\ld,\mu)=-\ld Y_B(\ld,\mu)$
where $Y_B(\ld,\mu)$ is the notation used there.
\end{rmk}

\section{Vertex operators and the Adler--Shiota--van Moerbeke formula}

Let us consider the additional symmetries acting on the tau function
of the two-component BKP hierarchy.


First we explain Sato's B\"{a}cklund transformation on the space of
tau functions in terms of vertex operators. For $k\in\Zop$, denote
\[
p_k=2\frac{\pd}{\pd t_k}, \quad p_{-k}=k\,t_k, \quad
\hat{p}_k=2\frac{\pd}{\pd \hat{t}_k}, \quad
\hat{p}_{-k}=k\,\hat{t}_k.
\]
Clearly $[\dot{p}_k,\dot{p}_l]=2 k\dt_{k,-l}$ with $\dot{p}_k=p_k$
or $\hat{p}_k$. Introduce $\dot{p}(z)=\sum_{k\in\Zop}\dot{p}_k
z^{-k}/k$, and define vertex operators
\begin{equation}\label{XXh}
X(\ld,\mu)=:e^{p(\ld)-p(\mu)}:, \quad
\hat{X}(\ld,\mu)=:e^{\hat{p}(-\ld^{-1})-\hat{p}(-\mu^{-1})}:.
\end{equation}
Here ``$:\ :$'' stands for the normal-order product, which means
that $\dot{p}_{k\ge0}$ must be placed to the right of
$\dot{p}_{k<0}$.

If either $\hat{\bf{t}}$ or $\bf{t}$ is fixed, the two-component BKP
hierarchy is reduced to the usual BKP hierarchy. Hence by using
Sato's B\"{a}cklund transformation for the latter (see
\cite{DKJM-KPBKP}), the vertex operators $\dot{X}(\ld,\mu)$ provide
infinitesimal transformations on the space of tau functions of the
two-component BKP hierarchy. Namely, given any solution
$\tau(\bf{t},\hat{\bf{t}})$ of \eqref{bl2BKP}, the functions
$\tau({\bf t},\hat{{\bf t}})+\ep \dot{X}(\ld,\mu) \tau({\bf
t},\hat{{\bf t}})+O(\ep^2)$ also satisfy \eqref{bl2BKP} modulo terms
of order $O(\ep^2)$ as $\ep\to0$.

\begin{thm}[ASvM formula] \label{thm-ASvM}
For the two-component BKP hierarchy the following equalities hold
ture
\begin{equation}\label{ASvM}
\dot{X}(\ld,\mu)\dot{w}(z)=2\frac{\mu-\ld}{\mu+\ld}\dot{Y}(\ld,\mu)\dot{w}(z).
\end{equation}
Here the actions of $\dot{X}(\ld,\mu)$ on the wave functions are
generated by their actions on the tau function (recall
\eqref{wtau}).
\end{thm}
\begin{prf}
The proof is similar to that for the case of the BKP hierarchy
\cite{Tu}, so we only sketch the main steps.

First, the bilinear equation \eqref{bl2BKP} yields the following Fay
identity
\begin{align}\label{}
&\frac{s_0-s_1}{s_0+s_1}\frac{s_0-s_2}{s_0+s_2}\frac{s_0-s_3}{s_0+s_3}\times
\nn \\
&\quad\times \tau({\bf
t}+2[s_0]+2[s_1]+2[s_2]+2[s_3],\hat{\bf{t}}+2[\hat{s}_1]+2[\hat{s}_2]+2[\hat{s}_3])
\tau({\bf t},\hat{\bf{t}}+2[\hat{s}_0]) \nn\\
&+\sum_{\mathrm{c.p.}(s_1,s_2,s_3)}
\frac{s_1-s_0}{s_1+s_0}\frac{s_1+s_2}{s_1-s_2}\frac{s_1+s_3}{s_1-s_3}\times
\nn \\
&\quad\times \tau({\bf
t}+2[s_2]+2[s_3],\hat{\bf{t}}+2[\hat{s}_1]+2[\hat{s}_2]+2[\hat{s}_3])
\tau({\bf t}+2[s_0]+2[s_1],\hat{\bf{t}}+2[\hat{s}_0]) \nn\\
=& \frac{\hat{s}_0-\hat{s}_1}{\hat{s}_0+\hat{s}_1}
\frac{\hat{s}_0-\hat{s}_2}{\hat{s}_0+\hat{s}_2}
\frac{\hat{s}_0-\hat{s}_3}{\hat{s}_0+\hat{s}_3}\times
\nn \\
&\quad\times \tau({\bf
t}+2[s_1]+2[s_2]+2[s_3],\hat{\bf{t}}+2[\hat{s}_0]+2[\hat{s}_1]+2[\hat{s}_2]+2[\hat{s}_3])
\tau({\bf t}+2[s_0],\hat{\bf{t}}) \nn\\
&+\sum_{\mathrm{c.p.}(\hat{s}_1,\hat{s}_2,\hat{s}_3)}
\frac{\hat{s}_1-\hat{s}_0}{\hat{s}_1+\hat{s}_0}
\frac{\hat{s}_1+\hat{s}_2}{\hat{s}_1-\hat{s}_2}\frac{\hat{s}_1+\hat{s}_3}{\hat{s}_1-\hat{s}_3}\times
\nn \\
&\quad\times \tau({\bf
t}+2[s_1]+2[s_2]+2[s_3],\hat{\bf{t}}+2[\hat{s}_2]+2[\hat{s}_3])
\tau({\bf t}+2[s_0],\hat{\bf{t}}+2[\hat{s}_0]+2[\hat{s}_1]).
\label{Fay}
\end{align}
Here $s_i$ and $\hat{s}_i$ are parameters, and ``c.p.'' stands for
``cyclic permutation''.

Second, take $\hat{s}_0=\hat{s}_3$ and $\hat{s}_1=-\hat{s}_2$, and
then let all $\hat{s}_i\to 0$. In this way \eqref{Fay} is reduced to
a Fay identity of the form for the BKP hierarchy. Differentiate it
with respect to $s_3$ and then let $s_3=0$, consequently one obtains
a differential Fay identity as follows
\begin{align}
&\left(\frac1{s_1^2}-\frac1{s_2^2}\right)\left(\tau({\mathbf
t}+2[s_1], \hat{\mathbf t})\tau({\mathbf t}+2[s_2], \hat{\mathbf t})
-\tau({\mathbf t}+2[s_1]+2[s_2], \hat{\mathbf t})\tau({\mathbf t},
\hat{\mathbf t})\right)
\nn \\
=&\left(\frac1{s_1}+\frac1{s_2}\right)\left(\pd_x\tau({\mathbf
t}+2[s_1], \hat{\mathbf t})\cdot\tau({\mathbf t}+2[s_2],
\hat{\mathbf t})-\tau({\mathbf t}+2[s_1], \hat{\mathbf
t})\cdot\pd_x\tau({\mathbf t}+2[s_2], \hat{\mathbf t})\right)
\nn \\
&-\left(\frac1{s_1}-\frac1{s_2}\right)\left(\pd_x\tau({\mathbf
t}+2[s_1]+2[s_2], \hat{\mathbf t})\cdot\tau({\mathbf t},
\hat{\mathbf t})-\tau({\mathbf t}+2[s_1]+2[s_2], \hat{\mathbf
t})\cdot\pd_x\tau({\mathbf t}, \hat{\mathbf t})\right).
\label{dFay1}
\end{align}
Exchange $s_i$ and $\hat{s}_i$ and repeat the above procedure, then
one gets another identity
\begin{align}
&\left(\frac1{\hat{s}_1^2}-\frac1{\hat{s}_2^2}\right)\left(\tau({\mathbf
t}, \hat{\mathbf t}+2[\hat{s}_1])\tau({\mathbf t}, \hat{\mathbf
t}+2[\hat{s}_2]) -\tau({\mathbf t}, \hat{\mathbf
t}+2[\hat{s}_1]+2[\hat{s}_2])\tau({\mathbf t}, \hat{\mathbf
t})\right)
\nn \\
=&\left(\frac1{\hat{s}_1}+\frac1{\hat{s}_2}\right)\left(\pd_{\hat{x}}\tau({\mathbf
t}, \hat{\mathbf t}+2[\hat{s}_1])\cdot\tau({\mathbf t}, \hat{\mathbf
t}+2[\hat{s}_2])-\tau({\mathbf t}, \hat{\mathbf
t}+2[\hat{s}_1])\cdot\pd_{\hat{x}}\tau({\mathbf t}, \hat{\mathbf
t}+2[\hat{s}_2])\right)
\nn \\
&-\left(\frac1{\hat{s}_1}-\frac1{\hat{s}_2}\right)\left(\pd_{\hat{x}}\tau({\mathbf
t}, \hat{\mathbf t}+2[\hat{s}_1]+2[\hat{s}_2])\cdot\tau({\mathbf t},
\hat{\mathbf t})-\tau({\mathbf t}, \hat{\mathbf
t}+2[\hat{s}_1]+2[\hat{s}_2])\cdot\pd_{\hat{x}}\tau({\mathbf t},
\hat{\mathbf t})\right). \label{dFay2}
\end{align}
Observe that \eqref{dFay1} and \eqref{dFay2} were also derived in
\cite{Ta-Fay} as the first and the second differential Fay
identities for the two-component BKP hierarchy.

Finally, by using the two differential Fay identities and
Lemma~\ref{thm-Yw}, the equalities \eqref{ASvM} are verified after a
straightforward calculation. Therefore the theorem is proved.
\end{prf}

We want to represent the additional symmetries \eqref{sml} and
\eqref{shml} with the tau function in \eqref{tau-1}. To this end we
expand the vertex operators in \eqref{XXh} formally as (cf.
\eqref{Y} and \eqref{Yh})
\begin{equation}\label{}
\dot{X}(\ld,
\mu)=\sum_{m=0}^\infty\frac{(\mu-\ld)^m}{m!}\sum_{l=-\infty}^\infty
\ld^{-m-l}\dot{W}_l^{(m)},
\end{equation}
where
\[
\dot{W}_l^{(m)}=\res_\ld\left(\ld^{m+l-1}\pd_\mu^m|_{\mu=\ld}\dot{X}(\ld,
\mu)\right).
\]
For convenience we assume $\dot{p}_i=0$ for even $i$. It is easy to
see
\begin{align*}
&W_l^{(0)}=\dt_{l,0}=\hat{W}_l^{(0)}, \quad W_l^{(1)}=p_l, \quad
\hat{W}_l^{(1)}=\hat{p}_{-l}, \\
&W_l^{(2)}=\sum_{i+j=l}:p_i p_j:-(l+1)p_l, \quad
\hat{W}_l^{(2)}=\sum_{i+j=-l}:\hat{p}_i
\hat{p}_j:-(l+1)\hat{p}_{-l}.
\end{align*}
Introduce two operators $Z(\ld,\mu)$ and $\hat{Z}(\ld,\mu)$ as
\begin{equation}\label{}
\dot{Z}(\ld,\mu)=\frac{1}{2}\frac{\mu+\ld}{\mu-\ld}(\dot{X}(\ld,\mu)-1).
\end{equation}
They are expanded to the form
\begin{equation}\label{}
\dot{Z}(\ld,\mu)=\sum_{m=0}^\infty\frac{(\mu-\ld)^m}{m!}\sum_{l=-\infty}^\infty
\ld^{-m-l}\dot{Z}_l^{(m+1)}
\end{equation}
with
\begin{equation}\label{}
\dot{Z}_l^{(1)}=\dot{W}_l^{(1)}, \quad
\dot{Z}_l^{(m+1)}=\frac{1}{m+1}\dot{W}_l^{(m+1)}+\frac{1}{2}\dot{W}_l^{(m)}
\hbox{ for } m\ge1.
\end{equation}

Now we have a corollary of Theorem~\ref{thm-ASvM}.
\begin{cor}\label{thm-taus}
For the two-component BKP hierarchy, the additional symmetries
\eqref{sml} and \eqref{shml} can be expressed via the tau function
as
\begin{equation}\label{taussh}
\frac{\pd\tau}{\pd
s_{m,m+l}}=\left(Z_l^{(m+1)}+\dt_{l0}c_m\right)\tau, \quad
\frac{\pd\tau}{\pd
\hat{s}_{m,m+l}}=\left(\hat{Z}_l^{(m+1)}+\dt_{l0}\hat{c}_m\right)\tau
\end{equation}
with certain constants $c_m$ and $\hat{c}_m$.
\end{cor}
\begin{prf}
Denote
\[
G({\bf
t};z)=\exp\left(-\sum_{k\in\Zop}\frac{2}{k\,z^k}\frac{\pd}{\pd t_k}
\right).
\]
By substituting the following equalities
\begin{align*}
&-(A_{m,m+l})_-w(z)=\frac{\pd w(z)}{\pd
s_{m,m+l}}=w(z)(G({\bf t};z)-1)\frac{\pd\tau/\pd s_{m,m+l}}{\tau}, \\
&\frac{1}{2}\frac{\mu+\ld}{\mu-\ld}X(\ld,\mu)w(z)=w(z)(G({\bf
t};z)-1)\frac{Z(\ld,\mu)\tau}{\tau}, \\
&(\hat{A}_{m,m+l})_+\hat{w}(z)=\frac{\pd \hat{w}(z)}{\pd
\hat{s}_{m,m+l}}=\hat{w}(z)(G(\hat{\bf t};-z^{-1})-1)\frac{\pd\tau/\pd \hat{s}_{m,m+l}}{\tau}, \\
&\frac{1}{2}\frac{\mu+\ld}{\mu-\ld}\hat{X}(\ld,\mu)\hat{w}(z)=\hat{w}(z)(G(\hat{\bf
t};-z^{-1})-1)\frac{\hat{Z}(\ld,\mu)\tau}{\tau}
\end{align*}
to the ASvM formula \eqref{ASvM}, one has
\[
\frac{\pd\tau}{\pd s_{m,m+l}}=(Z_l^{(m+1)}+c_{m,m+l})\tau, \quad
\frac{\pd\tau}{\pd
\hat{s}_{m,m+l}}=(\hat{Z}_l^{(m+1)}+\hat{c}_{m,m+l})\tau
\]
with constants $c_{m,m+l}$ and $\hat{c}_{m,m+l}$. They are recast to
the form \eqref{taussh} after a scaling transformation
$\tau\mapsto\tau\exp\left(-\sum_{m\ge0, l\ne m}(c_{m l} s_{m
l}+\hat{c}_{m l}\hat{s}_{m l})\right)$, which is allowed in the
definition of the tau function \eqref{tau-1}. The corollary is
proved.
\end{prf}


The representation \eqref{taussh} shows that the vector fields
$\pd/\pd s_{m l}$ and $\pd/\pd\hat{s}_{m l}$, when acting on the tau
function, generate a $W_{1+\infty}^B\times W_{1+\infty}^B$-algebra.
Here $W_{1+\infty}^B$ is the central extension of $w_{\infty}^B$,
see Proposition~\ref{thm-ssh}. It was observed by Adler, Shiota and
van Moerbeke \cite{ASvM} the phenomenon that lifting the action of
additional symmetries on wave functions to the action on tau
functions implies a central extension of $w$-algebras.

\section{Virasoro symmetries for Drinfeld--Sokolov hierarchies of type D}
\label{sec-virD}

Given an integer $n\ge3$, assume the two-component BKP hierarchy
\eqref{PPht} is constrained by $L^{2n-2}=\hat{L}^2=\cL$, where
\begin{equation}\label{mL}
   \cL=D^{2n-2}+\frac1{2}\sum_{i=1}^{n-1} D^{-1}\left(v_i
D^{2i-1}+D^{2i-1} v_i\right) +D^{-1} \rho D^{-1} \rho.
\end{equation}
Then the equations in \eqref{PPht} are reduced to
\begin{equation}\label{Dn}
\frac{\pd \cL}{\pd t_k}=[(L^k)_+, \cL], \quad \frac{\pd \cL}{\pd
\hat{t}_k}=[-(\hat{L}^k)_-, \cL], \qquad k\in\Zop.
\end{equation}
These equations compose the Drinfeld--Sokolov hierarchy of type
$D_n$, which is associated to the affine algebra $D_n^{(1)}$ and the
zeroth vertex of its Dynkin diagram \cite{DS, LWZ}. Note that the
equations \eqref{Dn} can also be written as \eqref{ppt1} and
\eqref{ppt2} via the dressing operators.

Generally, the additional symmetries \eqref{sml} and \eqref{shml}
for the two-component BKP hierarchy do not admit the constraint
$L^{2n-2}=\hat{L}^2$, hence they cannot be reduced to symmetries for
the subhierarchy \eqref{Dn} directly. However, a reduction works for
certain linear combinations of these symmetries.

For $j\ge-1$, let
\begin{equation}\label{}
B_j=\frac{1}{4n-4}A_{1,(2n-2)j+1}+\frac{1}{4}\hat{A}_{1,-2j+1}.
\end{equation}
Namely,
\begin{align*}
B_j=&\frac{1}{4n-4}(M L^{(2n-2)j+1}+ L^{(2n-2)j}M
L)+\frac{1}{4}(\hat{M} \hat{L}^{2j-1}+ \hat{L}^{2j}\hat{M}
\hat{L}^{-1}).
\end{align*}
Note that the product between $B_j$ (which may not lie in
$\cD^-\cup\cD^+$, see Section~2) and the bounded operator
$\cL\in\cD^-\cap\cD^+$ makes sense \cite{LWZ}. In fact, we have
\begin{align}\label{}
[B_j,\cL]=&-L^{(2n-2)(j+1)}+\hat{L}^{2(j+1)}=0. \label{BL}
\end{align}

\begin{rmk}
The last equality in \eqref{BL} is not true whenever $j<-1$, for the
reason that the operator $\cL$ has different inverses in $\cD^-$ and
$\cD^+$.
\end{rmk}

The equality \eqref{BL} together with the property $B_j^*=-D B_j
D^{-1}$ implies that the following evolutionary equations are well
defined
\begin{equation}\label{Ls}
\frac{\pd \cL}{\pd s_j}=[-(B_j)_-, \cL]=[(B_j)_+, \cL], \quad
j\ge-1.
\end{equation}
Since the operator $\cL$ can be written as $\cL=\Phi D^{2n-2}
\Phi^{-1}=\hat{\Phi} D^{-2} \hat{\Phi}^{-1}$ with dressing operators
of the form \eqref{Phi} and \eqref{phipsi}, then the equations
\eqref{Ls} can be redefined by either of the following
\begin{equation}\label{}
\frac{\pd \Phi}{\pd s_j}=-(B_j)_-\Phi, \quad
\frac{\pd\hat{\Phi}}{\pd s_j}=(B_j)_+\hat{\Phi}.
\end{equation}

Observe that $\pd/\pd s_j$ is just the reduction of the combination
\[
\frac{1}{4n-4}\frac{\pd}{\pd
s_{1,(2n-2)j+1}}+\frac{1}{4}\frac{\pd}{\pd \hat{s}_{1,-2j+1}}
\]
of the additional symmetries \eqref{sml} and \eqref{shml} for the
two-component BKP hierarchy. This observation leads to the following
consequence.
\begin{prp}
For the Drinfeld--Sokolov hierarchy of type $D_n$ defined by
\eqref{Dn}, the following statements hold true.
\\
(1) The flows \eqref{Ls} give symmetries of the hierarchy, namely,
\begin{equation*}
\left[\frac{\pd}{\pd s_j}, \frac{\pd}{\pd t_k}\right]=0, \quad
\left[\frac{\pd}{\pd s_j}, \frac{\pd}{\pd\hat{t}_k}\right]=0, \qquad
j\ge-1, ~k\in\Zop.
\end{equation*}
\\
(2) Let $w(z)$ and $\hat{w}(z)$ be the wave functions defined as in
\eqref{wavef}, then
\begin{equation}\label{waves}
\frac{\pd w(z)}{\pd s_j}=-(B_j)_-w(z), \quad
\frac{\pd\hat{w}(z)}{\pd s_j}=(B_j)_+\hat{w}(z), \qquad j\ge-1.
\end{equation}
\\
(3) With a tau function $\tau$ reduced from that of the
two-component BKP hierarchy, the above symmetries can be written as
\begin{equation}\label{tausj}
\frac{\pd\tau}{\pd s_j}=V_j\tau, \quad j\ge-1.
\end{equation}
Here the operators $V_j$ read
\begin{align}\label{Vj}
V_j=&\frac{1}{8n-8}\sum_i:p_i
p_{(2n-2)j-i}:+\frac{1}{8}\sum_i:\hat{p}_i
\hat{p}_{2j-i}:+\dt_{j0}\frac{n}{24}\left(1+\frac1{2n-2}\right),
\end{align}
and they obey the following commutation relation
\begin{equation}\label{ViVj}
[V_i, V_j]=(i-j)V_{i+j}+\dt_{i+j,0}\frac{n}{12}(i^3-i).
\end{equation}
\end{prp}
\begin{prf}
The first two assertions are clear. For the third item, according to
Corollary~\ref{thm-taus}, we see that the equalities \eqref{tausj}
hold for
\[
V_j=\frac{1}{4n-4}Z^{(2)}_{(2n-2)j}+\frac{1}{4}\hat{Z}^{(2)}_{-2j}+\dt_{j0}\cdot
\mathrm{const}.
\]
These operators are expanded to \eqref{Vj} with the constant chosen
appropriately (the chosen constant will be interpreted by the
spectrum of a Frobenius manifold, see the following section). The
commutation relation \eqref{ViVj} is checked by a straightforward
calculation, which is valid even for all $i,j\in\Z$, see, for
example, \cite{KR}. The proposition is proved.
\end{prf}

\section{Hierarchy associated to Frobenius manifold}

The hierarchy \eqref{Dn} carries a bi-Hamiltonian structure given by
the following compatible Poisson brackets (see Proposition 8.3 of
\cite{DS}, as well as \cite{DLZ, Wu}):
\begin{align}\label{poi1}
 \{F, G \}_1(\cL')&=2\int\res\, X\!\left( ( D Y_+\cL')_- - (\cL' Y_+  D)_-
+ (\cL' Y_- D)_+ - ( D Y_-\cL')_+ \right)\od x, \\
\{F, G \}_2(\cL')&=2\int\res\,\left(
 (\cL' Y)_+ \cL' X - X \cL'(Y \cL')_+ \right)\od x, \label{poi2}
\end{align}
in which  $\cL'=D\cL$, and the arbitrary local functionals $F$ and
$G$ have gradients $X={\dt F}/{\dt\cL'}$ and $Y={\dt G}/{\dt \cL'}$
respectively. More precisely, one rescales the time variables as
\begin{equation}\label{Tt}
T^{\al,p}=\left\{
\begin{array}{cl}
\dfrac{(2n-2)\Gm(p+1+\frac{2\al-1}{2n-2})}{\Gm(\frac{2\al-1}{2n-2})}t_{(2n-2)p+2\al-1},
\quad & \al=1, \dots, n-1; \\ \\
\dfrac{2\Gm(p+1+\frac12)}{\Gm(\frac12)}\hat{t}_{2p+1}, & \al=n
\end{array}\right.
\end{equation}
with $p=0, 1, 2, \dots$. In particular, $T^{1,0}=t_1=x$. Then the
hierarchy \eqref{Dn} is recast to the following bi-Hamiltonian form
\begin{equation}\label{Dnbh}
\frac{\pd F}{\pd T^{\al, p}}=\{F, H_{\al,
p}\}_1=\left(p+\frac12+\mu_\al\right)^{-1}\{F, H_{\al, p-1}\}_2,
\end{equation}
where the densities of the Hamiltonians $H_{\al, p-1}$ read
\[
h_{\al,p-1}=\left\{
\begin{array}{cl}
\dfrac{\Gm(\frac{2\al-1}{2n-2})}{(4n-4)\,\Gm(p+1+\frac{2\al-1}{2n-2})}\res\,L^{(2n-2)p+2\al-1},
\quad  & \al=1, \dots, n-1; \\ \\
\dfrac{\Gm(\frac12)}{4\,\Gm(p+1+\frac12)}\res\,\hat{L}^{2p+1}, &
\al=n,
\end{array}\right.
\]
and the constants $\mu_\al$ are
\begin{equation}\label{mual}
\mu_\al=\left\{
\begin{array}{cl}
\dfrac{2\al-n}{2n-2}, \quad & \al=1, \dots, n-1; \\
0, & \al=n.
\end{array}\right.
\end{equation}

Introduce a family of differential polynomials
\[
\Omega_{\al,p;\beta,q}=\pd_x^{-1}\frac{\pd h_{\al, p-1}}{\pd
T^{\beta,q}}, \quad \al, \beta\in\{1, 2, \dots, n\}; ~~p, q\ge0.
\]
They satisfy $\Omega_{\al,p;\beta,q}= \Omega_{\beta,q;\al,p}$, which
means that the densities $h_{\al,p}$ are tau-symmetric \cite{DZ}.
Hence a tau function $\tau$ of the integrable hierarchy \eqref{Dnbh}
is defined by
\begin{equation}\label{tauT}
\frac{\pd^2 \log \tau}{\pd T^{\al,p}\pd
T^{\beta,q}}=\Omega_{\al,p;\beta,q}.
\end{equation}
It is straightforward to see the following
\begin{lem}
For the hierarchy \eqref{Dn}, the tau function given in \eqref{tauT}
coincides with the one \eqref{tau-1} reduced from that of the
two-component BKP hierarchy.
\end{lem}

The dispersionless limit\,\footnote{According to the notation of
\cite{DZ}, one still needs to do a replacement $T^{\al,p}\mapsto\ep
T^{\al,p}$ with a small parameter $\ep$, meanwhile the residue of a
pseudo-differential operator becomes the coefficient of $(\ep
D)^{-1}$. By dispersionless limit we mean the limit as $\ep\to0$. }
of the Hamiltonian structures \eqref{poi1} and \eqref{poi2} are
associated to a Frobenius manifold on the orbit space of the Coxeter
group $D_n$ \cite{Du, DLZ, WX2}. Let $\cM$ denote this Frobenius
manifold.

We lay out some necessary datas of $\cM$. Take the symbol of the
pseudo-differential operator $\cL$ in \eqref{mL} as
\[
l(z)=z^{2n-2}+v_{n-1}z^{2n-4}+\dots+v_2 z^2+v_1+\rho^2 z^{-2}.
\]
Let
\begin{equation}\label{flatw}
w^\al=\left\{\begin{aligned}
    & -\frac{2n-2}{2n-2\al-1}\res_{z=\infty}~l(z)^{\frac{2n-2\al-1}{2n-2}}\od z,
     && \al=1,2,\cdots,n-1; \\
     & 2\,\rho,
     && \al=n,
     \end{aligned}
\right.
\end{equation}
then $w^1, \dots, w^n$ give a system of flat coordinates on $\cM$.
The potential $F(w^1, \dots, w^n)$ of this Frobenius manifold is
determined by
\begin{equation*}\label{}
\frac{\pd^3 F}{\pd w^\al \pd w^\beta \pd
w^\gm}=\frac1{2}\,\res_{l'(z)=0}\frac{\pd_{w^\al}
l(z)\cdot\pd_{w^\beta} l(z)\cdot\pd_{w^\gm} l(z)}{l'(z)} \od z.
\end{equation*}
On $\cM$ the unity vector field is $e=\pd/\pd w^1$, the invariant
metric and the multiplication are given by
\begin{align*}
& <\frac{\pd}{\pd w^\al},\frac{\pd}{\pd
     w^{\beta}}>=\eta_{\al\beta}, \quad
     \eta_{\al\beta}=\frac{\pd^3 F}{\pd w^1 \pd w^\al \pd w^\beta},
     \\
&\frac{\pd}{\pd w^{\al}}\cdot \frac{\pd}{\pd
w^{\beta}}=c^{\gm}_{\al\beta}\frac{\pd}{\pd
     w^{\gm}}, \quad
     c^{\gm}_{\al\beta}=\eta^{\gm\ep}\frac{\pd^3 F}{\pd w^\ep \pd w^\al \pd
     w^\beta}.
   \end{align*}
Here $(\eta^{\al\beta})=(\eta_{\al\beta})^{-1}$ and summations over
repeated Greek indices are assumed. The Euler vector field is
\[
E=\sum_{\al=1}^{n-1}\frac{n-\al}{n-1}w^\al\frac{\pd}{\pd w^\al}
   +\frac{n}{2n-2}w^n\frac{\pd}{\pd w^n},
\]
which satisfies $Lie_E F=(3-d)F$ with $d=(n-2)/(n-1)$.

It is easy to see that the dispersionless limit of $h_{\al,-1}$
equals to $\eta_{\al\beta}w^\beta$. Hence one has the following
lemma by virtue of the bi-Hamiltonian recursion relation.
\begin{lem}[see, for example, \cite{WX2}] \label{thm-ph}
The principal hierarchy \cite{DZ} for the Frobenius manifold $\cM$
coincides with the dispersionless limit of the hierarchy
\eqref{Dnbh}.
\end{lem}

By now we have explained that the hierarchy \eqref{Dnbh} satisfies
the axioms (A1) and (A2) reviewed in Section~1. From the principal
hierarchy to the full hierarchy, one still needs to consider the
Virasoro symmetries.

Recall that the spectrum of a Frobenius manifold is a quadruple
\begin{equation}\label{md}
(V,\,<\,,\,>,\,\mu,\,R),
\end{equation}
where $V$ is some linear space with a flat metric $<\,,\,>$ and
certain linear transformations $\mu$ and $R$. With this data, a
series of operators $\{L_j\mid j\in\Z\}$ were constructed in
\cite{DZ-vir, DZ}, which satisfy the Virasoro commutation relation
\begin{equation}\label{LiLj}
[L_i, L_j]=(i-j)L_{i+j}+\dt_{i+j,0}\frac{n}{12}(i^3-i).
\end{equation}

\begin{lem} \label{thm-LV}
For all $j\ge-1$, the Virasoro operators $L_j$ associated to the
Frobenius manifold $\cM$ coincide with $V_j$ in \eqref{Vj} derived
from vertex operators. Hence the symmetries \eqref{tausj} are
written as
\begin{equation}\label{tauL}
\frac{\pd\tau}{\pd s_j}=L_j\tau, \quad j\ge-1.
\end{equation}
\end{lem}
\begin{prf}
To compute the operators $L_j$ for the Frobenius manifold $\cM$, we
substitute into the formulae (2.29)--(2.32) of \cite{DZ-vir} with
the following data
\begin{itemize}\setlength{\itemsep}{0ex}
\item the metric $<\,,\,>$ given by $\eta=(\eta_{\al\beta})$ with
\[
\eta_{\al\beta}=\frac{\pd^3
F}{\pd{w^1}\pd{w^\al}\pd{w^\beta}}=\frac{1}{4n-4}\dt_{\al+\beta,n}+
\frac{1}{4}\dt_{\al,n}\dt_{\beta,n};
\]
\item $\mu=\dfrac{2-d}{2}\mathbf{1} -\nabla E= \diag(\mu_1,\mu_2,\dots,\mu_n)$ given
in \eqref{mual}, which satisfies
\[
\eta \mu+\mu \eta=0;
\]
\item $R=0$ for the reason that none of the differences $\mu_\al-\mu_\beta$ is a nonzero
integer.
\end{itemize}
Hence we have
\begin{align}
L_{-1}=&\sum_{p\ge1}T^{\al,p}\frac{\pd}{\pd
T^{\al,p-1}}+\frac1{2}\eta_{\al\beta}T^{\al,0}T^{\beta,0},
\\
L_{0}=&\sum_{p\ge0}\left(p+\frac1{2}+\mu_\al\right)T^{\al,p}\frac{\pd}{\pd
T^{\al,p}}+\frac1{4}\tr\left(\frac1{4}-\mu^2\right), \label{L0}
\\
L_{1}=&\sum_{p\ge0}\left(p+\frac1{2}+\mu_\al\right)\left(p+\frac{3}{2}+\mu_\al\right)
T^{\al,p}\frac{\pd}{\pd T^{\al,p+1}}
\nn \\
&+\frac1{2}\eta^{\al\beta}\left(\frac1{2}+\mu_\al\right)\left(\frac1{2}+\mu_\beta\right)
\frac{\pd}{\pd T^{\al,0}}\frac{\pd}{\pd T^{\beta,0}},
\\
L_{2}=&\sum_{p\ge0}\left(p+\frac1{2}+\mu_\al\right)\left(p+\frac{3}{2}+\mu_\al\right)
\left(p+\frac{5}{2}+\mu_\al\right) T^{\al,p}\frac{\pd}{\pd
T^{\al,p+2}} \nn \\
&+\eta^{\al\beta}\left(\frac1{2}-\mu_\al\right)\left(\frac1{2}-\mu_\beta\right)
\left(\frac3{2}-\mu_\beta\right) \frac{\pd}{\pd
T^{\al,1}}\frac{\pd}{\pd T^{\beta,0}}.
\end{align}
By using \eqref{Tt} and after a straightforward calculation, we
obtain $L_j=V_j$ for $j=-1,0,1,2$ where $V_j$ are given in
\eqref{Vj}. In particular, the constant term in \eqref{L0} is equal
to the one in $V_0$. By virtue of the Virasoro commutation relations
\eqref{ViVj} and \eqref{LiLj}, the lemma is proved.
\end{prf}

Getting Lemmas~\ref{thm-ph} and~\ref{thm-LV} together, we arrive at
the following conclusion.
\begin{thm}\label{thm-DZD}
According to Dubrovin and Zhang's construction, the hierarchy
associated to the semisimple Frobenius manifold $\cM$ and satisfying
the axioms (A1)--(A3) coincides with the hierarchy \eqref{Dnbh},
that is, the Drinfeld--Sokolov hierarchy of type $D_n$.
\end{thm}

\begin{rmk}
Observe that the definition of the tau function in \eqref{tauT} and
the linearization of the Virasoro symmetries \eqref{tauL} are
invariant with respect to the following scaling transformation
\[
F \mapsto \ld F, \quad \frac{\pd}{\pd T^{\al,p}}\mapsto \sqrt{\ld}
\frac{\pd}{\pd T^{\al,p}}, \quad T^{\al,p} \mapsto
\frac{1}{\sqrt{\ld}} T^{\al,p},
\]
where $F$ is the potential of the Frobenius manifold $\cM$ and $\ld$
is an arbitrary nonzero parameter. Without considering the Virasoro
symmetries, one can rescale $F$ and $T^{\al,p}$ separately and
define different tau functions for the same hierarchy. That is why a
tau function $\hat{\tau}=\tau^2$ for the hierarchy \eqref{Dn} was
obtained in \cite{LWZ}. In this sense the linearization of Virasoro
symmetries helps to select a unique tau function.
\end{rmk}

We write the Virasoro operators $L_j$ more presicely as
$L_j(\pd/\pd\mathbf{T}; \mathbf{T})$ with
$\mathbf{T}=\{T^{\al,p}\}$. For $j\ge-1$, introduce
\begin{equation}\label{Lhat}
\tilde{L}_j=L_j\left(\frac{\pd}{\pd{\mathbf T}}; \tilde{{\mathbf
T}}\right), \quad \tilde{{\mathbf T}}=\left\{\tilde{T}^{\al,p}=
T^{\al,p}-\dt^{\al}_1\dt^p_1\right\}.
\end{equation}
Clearly the operators $\tilde{L}_j$ also satisfy the Virasoro
commutation relation as in \eqref{LiLj}. As an application of the
property of linearization of Virasoro symmetries, we have the
following theorem. \
\begin{thm}\label{thm-Vc}
Suppose an analytic solution $\tau$ of the hierarchy \eqref{Dnbh}
satisfies the string equation
\begin{equation}\label{str}
L_{-1}\tau=\frac{\pd\tau}{\pd x},
\end{equation}
then it admits the following Virasoro constraints
\begin{equation}\label{}
\tilde{L}_j\tau=0, \quad j\ge-1.
\end{equation}
\end{thm}
\begin{prf}
Equation \eqref{str} is just $\pd\tau/\pd s_{-1}=\pd\tau/\pd x$,
which is represented with the wave functions as
\[
\frac{\pd w(z)}{\pd s_{-1}}=(D-z)w(z)=-L_- w(z), \quad
\frac{\pd\hat{w}(z)}{\pd s_{-1}}=D\,\hat{w}(z).
\]
These equations together with \eqref{waves} imply $B_{-1}=L$.

Recall for $j\ge-1$,
\[
B_j=\frac{1}{2n-2}(M L^{(2n-2)j+1}+ (n-1)j
L^{(2n-2)j})+\frac{1}{2}(\hat{M} \tilde{L}^{2j-1}-j \tilde{L}^{2j}).
\]
It is easy to see
\[
B_j=B_{-1}\cL^{j+1}=L^{(2n-2)(j+1)+1}.
\]
Hence
\begin{equation}\label{Phist}
\frac{\pd\Phi}{\pd s_j}=-(B_j)_-\Phi=-(L^{(2n-2)(j+1)+1})_-\Phi=
\frac{\pd\Phi}{\pd t_{(2n-2)(j+1)+1}}.
\end{equation}
From \eqref{wavef} and \eqref{wtau} it follows that
$\res\,\Phi=-2\pd_x\log\tau$. Taking the residue of \eqref{Phist},
we have
\[
\pd_x\left(\frac{1}{\tau}\left(\frac{\pd\tau}{\pd
s_j}-\frac{\pd\tau}{\pd t_{(2n-2)(j+1)+1}} \right) \right)=0.
\]
We make use of \eqref{tauL} and \eqref{Tt} to write the left hand
side of the above equation as $\pd_x(\tilde{L}_j\tau/\tau)$, thus
\begin{equation}\label{}
\tilde{L}_j\tau=f_j\tau, \quad j\ge-1,
\end{equation}
where $f_j$ are functions of $\{T^{\al,p}\mid (\al,p)\ne(1,0)\}$. In
particular, the string equation \eqref{str} means $f_{-1}=0$; we
need to show $f_j=0$ for all $j\ge0$.

Now we apply the Virasoro commutation relation for the operators
$\tilde{L}_j$. Denote
$\Dt_1=\sum_{p\ge1}\tilde{T}^{\al,p}\cdot{\pd}/{\pd T^{\al,p-1}}$,
then
\[
0=\tilde{L}_{-1}\tau=[\tilde{L}_0,
\tilde{L}_{-1}]\tau=-f_0\tilde{L}_{-1}\tau-\Dt_1(f_0)\cdot\tau=-\Dt_1(f_0)\cdot\tau.
\]
It means that $f_0$ is a constant. Similarly, considering
\begin{align}\label{}
&[\tilde{L}_{1}, \tilde{L}_{-1}]\tau=2\tilde{L}_{0}\tau=2 f_0\tau, \\
&[\tilde{L}_{1}, \tilde{L}_{0}]\tau=\tilde{L}_{1}\tau=f_1\tau,
\end{align}
we have
\[
-\Dt_1(f_1)=2 f_0, \quad -\Dt_0(f_1)=f_1,
\]
where
$\Dt_0=\sum_{p\ge0}\left(p+\frac{1}{2}+\mu_\al\right)\tilde{T}^{\al,p}\cdot{\pd}/{\pd
T^{\al,p}}$. Thus
\[
f_1=0, \quad f_0=0.
\]
The case $j=2$ is similar, and the cases $j\ge3$ follow from a
simple induction. The theorem is proved.
\end{prf}

Theorem~\ref{thm-main} follows from the combination of
Theorems~\ref{thm-DZD} and \ref{thm-Vc}.

\section{Concluding remarks}

We have obtained the linearized Virasoro symmetries for the
Drinfeld--Sokolov hierarchy of type $D_n$ starting from the
additional symmetries of its universal hierarchy, i.e., the
two-component BKP hierarchy. Therefore we complete the proof that
the Drinfeld--Sokolov hierarchy of type $D_n$ satisfies the axioms
(A1)--(A3) in Section~1 and show that its tau function admits the
Virasoro constraints. A tau function of the KdV or the extended Toda
hierarchies that satisfies the Virasoro constraints plays an
important role in the intersection theory on moduli spaces \cite{Ko,
DZ3, Wi}. Besides these hierarchies, now the Drinfeld--Sokolov
hierarchy of type $D_n$ is involved in Dubrovin and Zhang's
construction \cite{DZ}. For the topological solution of this
hierarchy, however, it is unknown whether there is a geometric
illustration in moduli spaces or not. We plan to study it in the
future. We also hope that our method, which is distinct from that
used for the extended Toda hierarchy in \cite{DZ3}, would be applied
to study Virasoro constraints for other integrable hierarchies.



\vskip 0.5truecm \noindent{\bf Acknowledgments.} The author thanks
Boris Dubrovin, Si-Qi Liu and Youjin Zhang for helpful discussions
and comments. This work is supported by the Young SISSA Scientist
Grant ``FIMA.645''.

\end{document}